\documentclass[
    ,final            % use final for the camera ready runs
  ]
  {aipproc}

\layoutstyle{6x9}

\newcommand\mnras{MNRAS}
\newcommand\apj{ApJ}
\newcommand\apjs{ApJS}
\newcommand\apjl{ApJL}
\newcommand\araa{ARA\&A}

\begin{document}

\title{The Evolution of AGN in Groups and Clusters} 

%\classification{<Replace this text with PACS numbers; choose from this list:
%                \texttt{http://www.aip..org/pacs/index.html}>}

\author{Paul Martini}{
  address={Department of Astronomy and Center for Cosmology and Astroparticle Physics, \\ The Ohio State University, Columbus, OH, USA}
}

%\author{<author2>}{
%  address={<common address for author2 and author3>}
%}

%\author{<author3>}{
%  address={<common address for author2 and author3>}
%  ,altaddress={<author1 address>} % additional visiting address
%}

\begin{abstract}
The evolution of AGN in groups and clusters provides important information 
about how their black holes grow, the extent to which galaxies and black holes 
coevolve in dense environments, and has implications for feedback in the local 
universe and at the epoch of cluster assembly. I describe new observations 
and analysis that demonstrates that the AGN fraction in clusters increases by 
a factor of eight from the local universe to $z \sim 1$ and that this evolution 
is consistent with the evolution of star-forming galaxies in clusters. The 
cluster AGN fraction remains approximately an order of magnitude below the 
field AGN fraction over this entire range, while a preliminary analysis of 
groups indicates that they too undergo substantial evolution. 
%The past decade has illustrated the dramatic importance of AGN feedback 
%on the hot gas in groups and clusters. While these AGN are generally 
%associated with the brightest member of the group or cluster, AGN are also 
%present in other galaxies in these environments. I will present new results 
%on the distribution of AGN in local group and cluster galaxies, as well as 
%the evolution of AGN in dense environments out to high redshift. The 
%demographics of local group and cluster AGN shed new light on the origin of 
%nuclear activity in galaxies, as well as their ability to retain cold gas, 
%while their evolution provides important new information on the coevolution of 
%black holes and galaxies as a function of environment. 
\end{abstract}

\maketitle

%%%%%%%%%%%%%%%%%%%%%%%%%%%%%%%%%%%%%%%%%%%%
%% MAINMATTER
%%%%%%%%%%%%%%%%%%%%%%%%%%%%%%%%%%%%%%%%%%%%

\section{Introduction}

Many studies over the past decade have presented strong evidence for the 
coevolution of black holes and galaxies based on samples dominated by 
the low-density field \citep[e.g.][and references 
therein]{hopkins06a,silverman08b}. It is interesting to determine if similar 
coevolution between AGN and galaxies is present in groups and clusters because 
the physical processes that drive galaxy evolution, such as the available cold 
gas to fuel star formation and black hole growth, are substantially different 
from the field. In addition, AGN in groups and clusters at low-redshift appear 
to play the critical role in maintaining the temperature of their hot 
atmospheres \citep[e.g.][]{mcnamara07}, while AGN heating at the epoch of 
cluster assembly remains a viable explanation for the minimum entropy level 
in the intracluster medium. 

These questions have motivated my collaborators and I to systematically search 
for AGN in groups and clusters of galaxies and led to the discovery of large 
numbers of AGN in these dense environments in the local universe 
\cite{martini06,martini07}. In this contribution I summarize some recently 
published work on the evolution of AGN in clusters of galaxies 
\cite{martini09} and present a new, preliminary analysis of the evolution of 
AGN in the lower-density group environment. I conclude with a brief discussion 
of several future directions. 

\section{Methodology} 
 
The key requirement to measure the evolution of AGN in groups and clusters, as 
well as perform quantitative comparisons to their galaxy populations,
is systematic and unbiased selection. We select AGN with X-ray observations 
obtained by the {\it Chandra} and {\it XMM} satellites because X-ray emission 
provides a relatively unbiased measure of accretion for luminous objects. 
Redshifts of the counterparts to the X-ray sources, obtained from our work or 
the literature \citep{silverman05a,eckart06,martini07,eastman07,martini09}, are 
then used to identify the subset of X-ray sources associated with the groups 
and clusters\footnote{Distinct X-ray emission from the central galaxy is 
difficult to identify due to the presence of extended, hot gas emission. We 
consequently do not include these galaxies in our study.}. Our final sample 
includes 17 clusters with $z<0.4$ and 15 clusters with $0.4<z<1.3$ that are 
reasonably well-matched in spite of their heterogeneous selection based on 
available archival data. 

We systematically characterize the AGN population via the AGN fraction, which 
is defined as the fraction of all galaxies brighter than some absolute 
magnitude that host AGN above some luminosity threshold. Given the sensitivity 
limits of X-ray data and multi-object spectroscopy, the AGN fraction in 
clusters is defined here as the fraction of galaxies brighter than $M_R^* + 1$ 
that host AGN with hard X-ray [2-10 keV] luminosities above $10^{43}$ erg 
s$^{-1}$. While redshift measurements are nearly complete for the X-ray 
sources, this is not always the case for the entire galaxy population of the 
group or cluster. In many cases the number of galaxies above the 
luminosity threshold is estimated from the velocity dispersion and an 
empirical relation derived from the SDSS \citep{becker07}. Throughout this 
work the AGN fraction is only measured for AGN and other galaxies that fall 
within the projected $r_{200}$ radius, that is the radius within which the 
group or cluster is a factor of 200 overdensity. Further details are provided 
in \cite{martini09}. 

\subsection{Results} 

%%%%%%%%%%%%%%%%%%%%%%%%%%%%%%%%%%%%%%%%%%%%
%% Sample figure:
%%
%% The option [height=...] scales the picture to the given height,
%% without it it would be printed at its nominal size
%%%%%%%%%%%%%%%%%%%%%%%%%%%%%%%%%%%%%%%%%%%%

The main result of this analysis is presented in the Figure, which demonstrates
that the cluster AGN fraction increases by a factor of eight from the present 
to $z \sim 1$. This evolution is based on 2 luminous AGN in 17 clusters at 
$z<0.4$ and 17 luminous AGN in 15 clusters at $z>0.4$. This dramatic evolution 
is similar to the evolution of the star-forming galaxy population in clusters 
known as the Butcher-Oemler effect \citep{butcher84}. Here we have 
parametrized the AGN evolution as $f_A \propto (1+z)^\alpha$ where 
$\alpha_{AGN} = 5.3^{+1.8}_{-1.7}$ ({\it dashed line} in the Figure). This 
power-law index is consistent with the value of 
$\alpha_{SF} = 5.7^{+2.1}_{-1.8}$ recently measured for star-forming 
galaxies from mid-infrared observations \citep{haines09} and supports the 
hypothesis that black holes and galaxies coevolve in dense environments. 

\begin{figure}
  \includegraphics[height=.5\textheight]{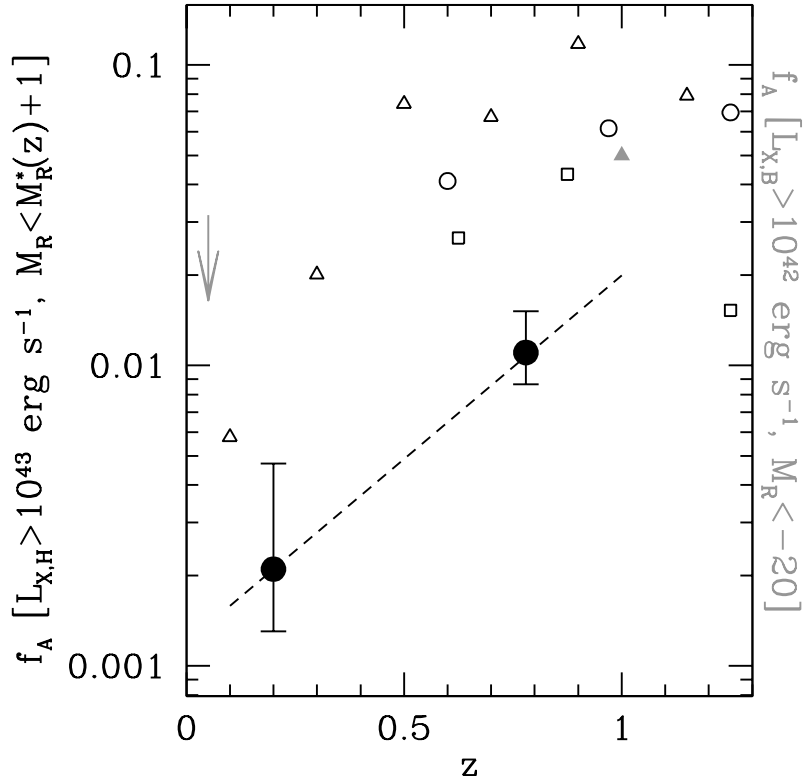}
  \caption{Evolution of the AGN population in clusters from $z = 0$ to
$z = 1.3$ from a sample 17 clusters at $z<0.4$ and 15 clusters at $z>0.4$ 
({\it filled circles}). These points represent the average fraction of cluster 
members more luminous than $M_R^* + 1$ that host AGN more luminous than 
$L_{X,H} \geq 10^{43}$ erg/s. The increase in the AGN fraction is 
consistent with a power-law of the form $f_A \propto (1+z)^{5.3}$ ({\it dashed 
line}) and is approximately an order of magnitude below the field AGN fraction 
({\it open symbols}) over this redshift range. A preliminary estimate of the 
group AGN fraction suggests their rate of evolution is similar ({\it gray 
triangle} at $z=1$ and {\it upper limit} at $z=0.05$). The group AGN fraction 
is defined as the fraction of $L_{X,B} \geq 10^{42}$ erg/s AGN 
in galaxies with $M_R < -20$ mag (see {\it right-hand axis} label). 
}
\end{figure}

The rate of evolution of the AGN fraction also appears similar to 
the lower-density field ({\it open symbols}), although the cluster AGN 
fraction is approximately an order-of-magnitude lower over the entire 
redshift range. The lower AGN fraction in clusters relative 
to the field strongly suggests that there must be a substantial change 
in the AGN fraction in the intermediate-density group environment. 
Observations of high-redshift groups find a substantial decrease in the AGN 
fraction with increasing overdensity at $z \sim 1$ \citep{silverman09b} 
and a measured group AGN fraction of $\sim 5$\% for AGN with broad-band
X-ray luminosities $L_{X,B} > 10^{42}$ erg s$^{-1}$ in galaxies more luminous 
than $M_R < -20$ mag \citep[][{\it gray triangle} in Figure]{georgakakis08b}. 
Studies of low-redshift groups find that the AGN fraction is somewhat higher 
but consistent with the value measured in clusters 
\citep{shen07,sivakoff08,arnold09}; however, the group sample contains far 
fewer galaxies and provides only an upper limit on AGN as luminous as those in 
the high-redshift sample (the {\it arrow} at $z=0.05$ in the Figure represents 
the 90\% confidence limit on the group AGN fraction).

%%%%%%%%%%%%%%%%%%%%%%%%%%%%%%%%%%%%%%%%%%%%
%% SAMPLE TABLE
%%
%% Shows the use of \tablehead and \tablenote
%% macros
%%%%%%%%%%%%%%%%%%%%%%%%%%%%%%%%%%%%%%%%%%%%

%\begin{table}
%\begin{tabular}{lrrrr}
%\hline
%  & \tablehead{1}{r}{b}{Single\\outlet}
%  & \tablehead{1}{r}{b}{Small\tablenote{2-9 retail outlets}\\multiple}
%  & \tablehead{1}{r}{b}{Large\\multiple}
%  & \tablehead{1}{r}{b}{Total}   \\
%\hline
%1982 & 98 & 129 & 620    & 847\\
%1987 & 138 & 176 & 1000  & 1314\\
%1991 & 173 & 248 & 1230  & 1651\\
%1998\tablenote{predicted} & 200 & 300 & 1500  & 2000\\
%\hline
%\end{tabular}
%\caption{Average turnover per shop: by type
%  of retail organisation}
%\label{tab:a}
%\end{table}

\section{Summary and Future Directions} 

The observations summarized here have shown that the AGN fraction in clusters 
increases by approximately an order of magnitude from the present to 
$z \sim 1$. This evolution is also quantitatively similar to the 
evolution of the star-forming galaxy fraction in clusters, which suggests 
that cluster galaxies and their central black holes also coevolve, although 
at a different rate from their counterparts in the field. A simple comparison 
between the AGN fraction in clusters and the field indicates that the 
cluster AGN fraction is approximately an order of magnitude lower over the 
entire observed redshift range. 

One future direction for this research is to push further into the past 
toward the epoch of cluster galaxy assembly at $z\sim2 - 3$. Numerous lines of 
evidence suggest that cluster galaxies formed earlier than their 
field counterparts, and the hypothesis of black hole and galaxy coevolution 
predicts that the the AGN fraction in dense environments should exceed the 
field value by this point. At lower redshifts, the identification of the 
properties of the ``transition'' group environment, in which the AGN fraction 
drops from the field to the cluster value, can potentially shed substantial 
light on the nature of the mechanism(s) responsible for triggering and 
fueling AGN. Finally, measurements of star formation and black hole accretion 
rates in the same cluster samples could reveal the extent of the correlation 
exhibited by individual galaxies and provide new insights into the physical 
processes that drive the observed coevolution. 

%%%%%%%%%%%%%%%%%%%%%%%%%%%%%%%%%%%%%%%%%%%%%%%%
%% BACKMATTER
%%%%%%%%%%%%%%%%%%%%%%%%%%%%%%%%%%%%%%%%%%%%%%%%

\begin{theacknowledgments}
I am grateful for the contributions of T. Arnold, A. Berti, T.~E. Jeltema, 
D.~D. Kelson, J.~S. Mulchaey, G.~R. Sivakoff, and A.~I. Zabludoff to the 
research that I have presented in this conference article. I also appreciate 
the opportunity to speak at the conference and thank the organizers for this 
stimulating meeting. My research was supported in part by NASA through Chandra 
Award Number AR8-9014X and Spitzer Grant 1364997, and NSF via award 
AST-0705170, and the Department of Astronomy at OSU. 

\end{theacknowledgments}

\end{document}